# Microwave Induced Zero-Conductance State in a Corbino Geometry 2D Electron Gas with Capacitive Contacts

A. A. Bykov, I. V. Marchishin, A. V. Goran, and D. V. Dmitriev

*Institute of Semiconductor Physics, 630090 Novosibirsk, Russia*

Microwave induced photoconductivity of a two-dimensional electron gas (2DEG) in selectively-doped GaAs/AlAs heterostructures has been studied using the Corbino geometry with capacitive contacts at a temperature $T$=1.6K and magnetic field $B$ up to 0.5 T. Zero-conductance states have been observed in the samples under study subject to microwave radiation, similarly to the samples with ohmic contacts. It has been shown that ohmic contacts do not play a significant role for observation of zero-conductance states induced by microwave radiation.

Giant magnetoresistance oscillations in a 2D electron system at high filling factor subject to microwave radiation were first observed about 10 years ago [1,2]. Till now these oscillations are a subject of many investigations [3,4]. There is still a discussion about the nature of zero-resistance and zero-conductance states in the minima of these oscillations [5-14]. One of the points of interest is the role of ohmic contacts to a 2D electron system on zero-resistance and zero-conductance states [15]. Both 2D electron gas and its ohmic contacts are subject to microwave radiation and this may lead not only to the rise of photo-EMF but also contribute to photoconductivity.

Microwave-induced giant magnetoresistance and magnetoconductance oscillations in a 2D electron system are periodic in inverse magnetic field with the maxima approximately at the harmonics of the cyclotron resonance. We can write the conductance of a 2D electronic gas subject to microwave radiation of frequency $\omega/2\pi$ as $G^\omega = G^0 + \Delta G^\omega$ where $G^0$ is the conductivity in the absence of radiation and $\Delta G^\omega$ is the microwave induced photoconductivity responsible for oscillations in $G^\omega(B)$. $\Delta G^\omega$ is a sign-alternating function of $j = \omega/\omega_c$ [20-25] where $w_c$ is the cyclotron frequency and $j = 1, 2, 3\ldots$ are positive integers. According to the theory developed in [20-25] the conductivity in the minima of giant oscillations is negative. However the experimental curves show the value of conductivity that is very close to zero, which is a subject of an active discussion [12-14].

Until now the microwave induced zero-conductance states in a 2D electron gas were only investigated using samples with ohmic contacts [8,11]. It has been recently shown that zero-conductance states are not observed when using a non-contact method based on the measurement of high-frequency signal damping along a coplanar waveguide positioned on the sample surface [15]. In this paper we used Corbino rings with capacitive contacts to measure the microwave-induced conductivity of a 2D electron gas. Zero-conductance states have been observed in such samples subject to microwave radiation, similarly to the samples with ohmic contacts [8,11]. Therefore it has been experimentally confirmed that ohmic contacts are not essential for observation and origination of zero-conductance states induced by microwave radiation in 2D electron systems at large filling factor.

The selectively doped heterostructures under study were single GaAs quantum wells with AlAs/GaAs superlattice barriers. The width of quantum wells was 13 nm, the distance between the well and sample surface was 105 nm. The structures were grown using molecular-beam epitaxy on (100) GaAs substrates. The concentration and mobility of heterostructures were calculated using the van der Pauw method on 5x5 mm square samples with ohmic contacts at the corners of a square [26]. Ohmic contacts were produced by burning indium. The sides of a square were oriented along [110] and [1$\bar{1}$0] directions that corresponds to the maximum and minimum electron mobility in GaAs/AlAs heterostructures [27,28]. The concentration of electrons was $n_e = 8\times10^{15}$ m$^{-2}$. The mobility $\mu_x$ and $\mu_y$ along the directions [110] and [1$\bar{1}$0] was calculated from $\rho_{xx}$ and $\rho_{yy}$ in zero magnetic field, $n_e$ was calculated from Hall resistance. At the temperature $T$ = 4.2 K the mobility was: $\mu_x$=170 m$^2$/Vs and $\mu_y$=210 m$^2$/Vs.

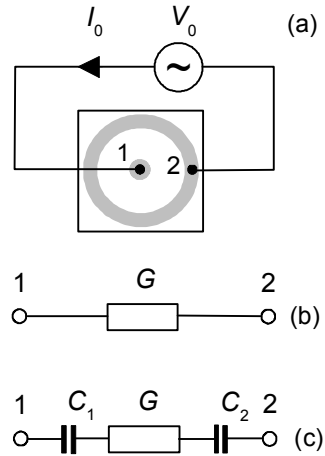

Fig. 1. (a) The conductivity measurement scheme in the Corbino geometry. Metallic electrodes 1 and 2 are in gray. (b) The equivalent scheme of a Corbino ring with ohmic contacts. (c) The equivalent scheme of a Corbino ring with capacitive contacts.

Fig. 1a shows the scheme of Corbino ring measurements. Originally all the samples were 5x5 mm squares, same as the one used for van der Pauw measurements. Two electrodes were placed at the planar surface of samples. The electrodes are shown as gray on Fig. 1a. The diameter of the inner electrode was ~0.5 mm and the distance between the inner and outer electrodes was ~1.5 mm. In the case of ohmic contacts the electrodes were burned, while for capacitive contacts they weren't. The *ac* current resistance between capacitive contacts and the 2D electron gas exceeded 40 M$\Omega$. The measurements were performed at $T = 4.2$ K in magnetic field $B$ up to 0.5 T and the measuring current frequency $f = 1$ MHz. The voltage $V_0$ between contacts in Corbino rings didn't exceed 0.1 mV. Lock-in technique was used for measurements of imaginary and real parts of $I_0$. We used a round waveguide with the inner diameter 6mm to guide microwave radiation of frequency $F = 145$ GHz to the samples. The samples were placed few millimeters from an open end of the waveguide. The maximum output power of radiation was $P_{out} \sim 4$ mW.

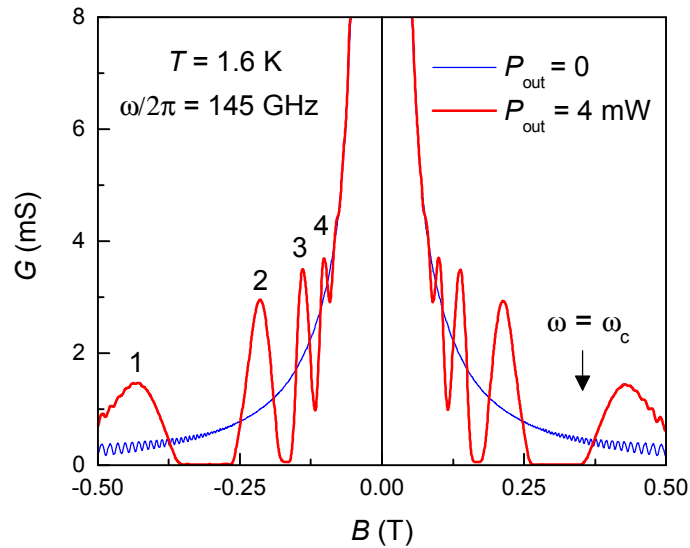

Fig. 2. $G(B)$ measured on a 2D Corbino ring with ohmic contacts with and without microwave radiation of frequency 145 GHz. Thin line corresponds to $P_{out} = 0$, thick line corresponds to $P_{out} = 4$ mW. The arrow marks the cyclotron resonance.

Fig. 1b shows the equivalent scheme of Corbino ring with ohmic contacts. $G$ denotes the conductivity of a 2D electron gas between ohmic contacts. The conductivity of ohmic contacts in the Corbino disks under study was much greater than $G$ so we didn't take it into account. Fig. 2 shows the experimental curves of conductivity with and without microwave radiation. The oscillating component appears in the presence of microwave radiation, its maxima are marked by numbers 1, 2, 3 and 4. The position of the maxima is determined by $j = \omega/\omega_c$. It's clearly seen that the conductivity between the maxima 1 and 2 is very close to zero, which is in agreement with results presented in [8, 11].

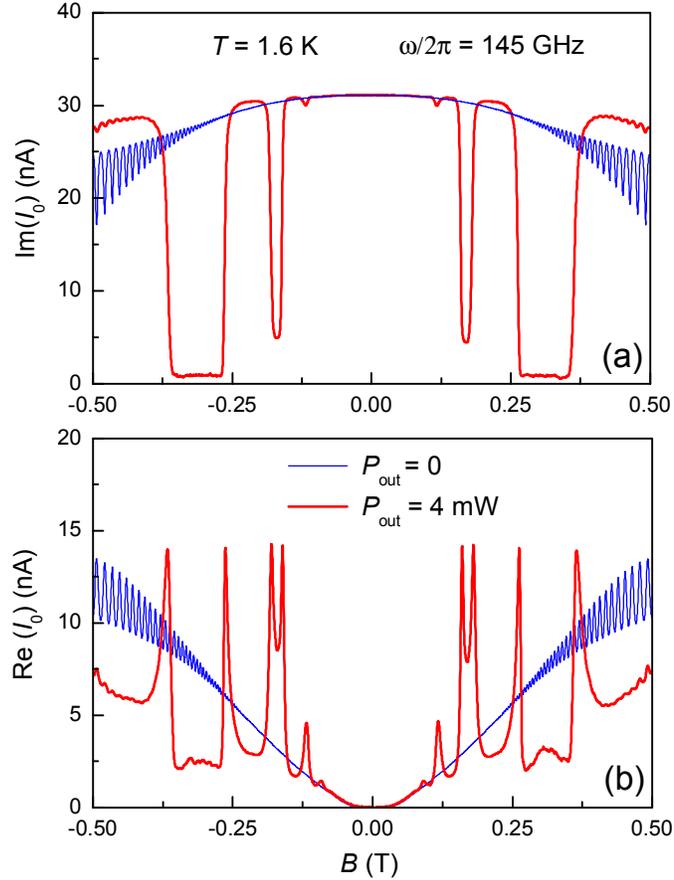

Fig. 3. (a) Im($I_0$) vs $B$ and (b) Re($I_0$) vs $B$ measured on a 2D Corbino ring with capacitive contacts with and without microwave radiation of frequency 145 GHz. Thin line corresponds to $P_{out} = 0$, thick line corresponds to $P_{out} = 4$ mW.

Fig. 1c shows the equivalent scheme of Corbino ring with capacitive contacts. The voltage $V_0$ is applied to capacitive contacts 1 and 2 with capacitances $C_1$ and $C_2$, respectively. We can replace these capacitances with an equivalent capacitance $C = C_1C_2/(C_1 + C_2)$ and taking into account that $C_1 \ll C_2$, we can write $C \approx C_1$. The electric current $I_0$ of the frequency $f$, going through a Corbino ring with capacitive contacts, can be written as:

$$I_0 = V_0/[1/G + 1/(i2\pi fC)]. \tag{1}$$

It's easy to show using (1) that $G$ can be written as:

$$G = [\text{Re}^2(I_0) + \text{Im}^2(I_0)]/[V_0 \text{Re}(I_0)]. \tag{2}$$

The capacitance $C$ in the disks under study depended on the diameter of the waveguide 1 and varied from 50 to 350 pF in different disks.

Fig. 3 shows the curves of imaginary and real parts of $I_0$ vs magnetic field $B$ measured in Corbino ring geometry with the capacitance $C \sim 50$ pF. The oscillating components of the imaginary and real parts of $I_0$ in the presence of microwave radiation can be clearly seen. Fig. 4 shows the curves of $G(B)$ calculated using (2). With no microwave radiation the value of $G$ decreases monotonically with the increase of magnetic field, while in the presence of radiation an oscillating component appears in $G(B)$. This component is fully identical to giant oscillations of magnetoconductivity presented on Fig. 2.

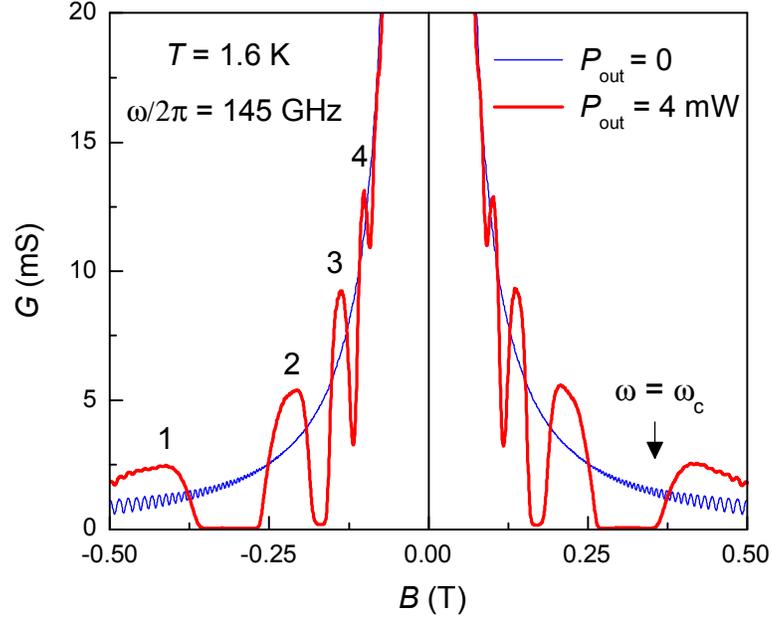

Fig. 4. $G(B)$ measured on a 2D Corbino ring with capacitive electrodes with and without microwave radiation of frequency 145 GHz. Thin line corresponds to $P_{out} = 0$, thick line corresponds to $P_{out} = 4$ mW. The arrow marks the cyclotron resonance.

To summarize, in this paper we have shown that microwave induced conductivity measured in a 2D electron gas at large filling factor in Corbino ring geometry is identical when using ohmic and capacitive contacts. It has been experimentally shown that ohmic contacts are not important for observation of microwave-induced zero-conductance states in 2D electron systems at large filling factor.

The work was supported by RFFI project #08-02-01051.